\documentclass[journal]{IEEEtran}
\usepackage{tikz}
\usepackage{xcolor,soul,framed} 
\usepackage{tabularx} 
\usepackage{cite}
\usepackage[flushleft]{threeparttable}
\usepackage{url}
\usepackage{graphicx}

\newcommand*\circled[1]{\tikz[baseline=(char.base)]{
            \node[shape=circle,draw,inner sep=1pt] (char) {\textcolor{black}{\textbf{#1}}};}}
            
\colorlet{shadecolor}{yellow}
\usepackage[cmex10]{amsmath}
\usepackage{array}
\usepackage{mdwmath}
\usepackage{mdwtab}
\usepackage{eqparbox}
\usepackage{url}
\usepackage{todonotes}
\usepackage{listings}    
\usepackage{subfig} 
\usepackage{hyperref}
\usepackage{etoolbox}

\usepackage{tabularx}
\usepackage{adjustbox}

\usepackage{tikz}

\usepackage{algorithm}
\usepackage{algorithmic}
\usepackage{multirow}
\usepackage{verbatimbox}
\usepackage{authblk}

\makeatletter
\newcommand\fs@norules{\def\@fs@cfont{\bfseries}\let\@fs@capt\floatc@ruled
  \def\@fs@pre{}%
  \def\@fs@post{}%
  \def\@fs@mid{\kern3pt}%
  \let\@fs@iftopcapt\iftrue}
\makeatother
\floatstyle{norules}
\restylefloat{algorithm}
\usepackage{fixltx2e}

\makeatletter
\patchcmd{\@verbatim}
  {\verbatim@font}
  {\verbatim@font\small}
  {}{}
\makeatother

\begin{document}
\bstctlcite{IEEEexample:BSTcontrol}
\title{Rewrite to Reinforce: Rewriting the Binary to Apply Countermeasures against Fault Injection}

\author[1]{Pantea Kiaei}
\author[2]{Cees-Bart Breunesse}
\author[3]{Mohsen Ahmadi}
\author[1]{Patrick Schaumont}
\author[2]{Jasper van Woudenberg}
\affil[1]{Worcester Polytechnic Institute}
\affil[2]{Riscure}
\affil[3]{Arizona State University}




\maketitle

\begin{abstract}
Fault injection attacks can cause errors in software for malicious purposes. Oftentimes, vulnerable points of a program are detected after its development. It is therefore critical for the user of the program to be able to apply last-minute security assurance to the executable file without having access to the source code. In this work, we explore two methodologies based on binary rewriting that aid in injecting countermeasures in the binary file. The first approach injects countermeasures by reassembling the disassembly whereas the second approach leverages a full translation to a high-level IR and lowering that back to the target architecture. 

\end{abstract}

\begin{IEEEkeywords}
Fault injection, binary protection, targeted countermeasures, automated countermeasure, binary rewriting
\end{IEEEkeywords}

%
\IEEEpeerreviewmaketitle


\section{Introduction}

\IEEEPARstart{N}{owadays}, the Fault Injection (FI) hardware attacks are becoming more 
prevalent. Successful fault attacks lead to information leakage \cite{piret2003differential}, \cite{leveugle2009statistical} or privilege escalation. 
While fault injection is targeted at the hardware (e.g., clock glitching), consequences of the resulting faults may affect the software running on a processor.
For example, ARM's secure boot can be affected by voltage glitching to enable an attacker to load controlled values in the program counter (PC) \cite{7774479}. Furthermore, Vasselle et al. \cite{8419264} show how laser injected faults can bypass the secure bootloader on an android smartphone.

To defend against these attacks, an extensive amount of countermeasures have been proposed that can be categorized in three groups \cite{belleville2018automatic}, namely those that can be applied to the software source code, those that are implemented within the compiler tool-chain, and those that are directly applied to the execution binary. The first two categories require access to the source code of the program which may not be practical in some scenarios. For instance, binary-level protection is useful for legacy binary code, or for third-party library code, or even for binary code for which the source code has been lost. In this paper, we target the problem of applying FI countermeasures when we do not have access to the source code of the program. Knowing that applying countermeasures directly to the binary file is not easy, we demonstrate how static binary rewriting approaches help 
with instrumenting the program with our countermeasures.
We apply and evaluate two static binary rewriting schemes; One reassembleable disassembly and the other complete translation.
Using the reassembleable disassembly method, we demonstrate how we can apply simple fixes to the binary file with low overhead. Using the full-translation to LLVM-IR approach, we show how more complex countermeasures can be implemented exploiting the power of an intermediate representation (IR).

The rest of the paper is organized as the following:
In Section~\ref{sec:related}, we discuss the related work. 
In Section~\ref{sec:binarylifter}, we give a brief background on binary rewriting.
In Section~\ref{sec:approaches}, we introduce our countermeasure insertion methodologies. 
In Section~\ref{sec:experiments}, we show the results of simple FI countermeasures implemented using our proposed approaches. Finally, in Section~\ref{sec:conclusion}, we conclude the paper. 

\section{Related Work}\label{sec:related}

\begin{figure*}[t]
    \centering
    \includegraphics[width=.8\linewidth]{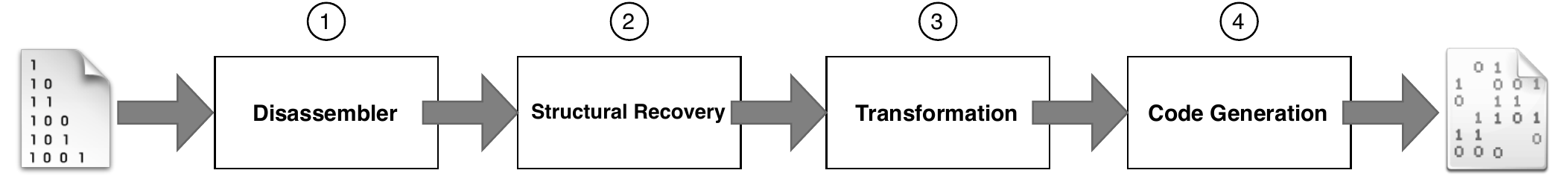}
    \caption{Coverage percentage achieved over for the cover sizes of different sizes\label{fig:pipeline}}
    \vspace{-.5cm}
\end{figure*}

While the bulk of fault countermeasures is based in detecting faults in redundant design, researchers have made many different proposals regarding the format and abstraction level of these redundancies. Some of the related work start from the source code and add the countermeasure at that high level of abstraction. 
For instance, Lalande et al. \cite{lalande2014software} proposed a counter-based approach in which a counter is incremented and checked after a set number of instructions to detect jump attacks. In other works, the countermeasures are added at compile time and require access to the source code. Barry et al. modified the LLVM back-end for insertion of countermeasures against instruction-skip fault attacks. For this purpose, they duplicated instructions based on whether the operation is idempotent \cite{barry2016compilation}.
Our focus is on scenarios where the source code of the program is unavailable and we need to protect the executable file against fault attacks. Given-Wilson et al. \cite{given2019automated} propose a methodology to detect vulnerabilities of program binaries to fault injection. 
In their methodology, they annotate the source code with safety properties and detect vulnerabilities when these properties are shown to not be held using model-checking.
Br{\'e}jon et al. \cite{brejon2019fault} propose a framework consisting of symbolic execution, static analysis, and model-checking to find vulnerabilities in the binary file. Both of these works focus on finding vulnerabilities but once the vulnerabilities are found, the source code should be accessible to add corresponding countermeasures. 
De Keulenaer et al. \cite{de2016link} use the link-time optimizer tool Diablo \cite{van2005diablo} and look for patterns of instructions in the assembly-level IR that are known to be vulnerable to fault injection and replace them with hardened code. 
They show the approach of using the assembly-level IR results in a more compact hardened code, compared to the compiler-level IR, which is important for small embedded systems. 
In another work, O'sullivan et al. \cite{o2011retrofitting} propose a lifting and rewriting methodology based on the SecondWrite tool \cite{anand2010decompilation} for hardening a binary file against low-level software attacks such as buffer overflow attacks. Fault injection attacks are not considered in their work.

In this project, we propose two different approaches to find the vulnerabilities in the binary code against fault injection attacks and add corresponding countermeasures. 
As the first contribution of this paper, we propose a simulation-driven countermeasure insertion. In this method, only the vulnerable parts of the binary file are patched and hence the overhead of the protected code is decreased compared to a full application of the countermeasure. As a second contribution, this work for the first time shows how lifting the binary to an IR can help in adding countermeasures against fault attacks. We use two different IR levels to this end and compare their resulting overhead.

\section{Binary Rewritering} 
\label{sec:binarylifter}

In this section, we will go through the publicly available binary rewriting solutions and compare their approach with regard to structural recovery, data type extraction, and limitations on supported architectures. 

\subsection{Definition} \label{sec:definiion} Binary rewriting is denoted as the process of modifying a compiled program in such a way that it remains executable and functional without having access to the source-code. There are two types of binary modifiers, static and dynamic. In the static approach, results of the modification will be stored on a persistent memory like disk for future execution. However, dynamic rewriting is applied during the program execution. In this paper we focus on static binary rewriting technique and compare the tools following this scheme.

Based on Figure \ref{fig:pipeline}, from a high-level view, in the first step (\circled{1}), a binary rewriter receives a file in a binary stream format as input and passes it to the disassembler for decoding the instructions, and retrieving global variables and sections. Decoded instructions help step \circled{2} in building the control and data flow, recovering data types, and function boundaries to semantically enriching the context with metadata lost during the compilation.
Transformation step (\circled{3}) modifies the target binary in a way that mutated output is a working executable.

\subsection{Static Binary Rewriting}
There are three known rewriting schemes. The oldest one is based on \textit{detouring} at assembly level. Detouring works by hooking out the underlying instruction. There are two flavors of the detouring technique, \textit{patch-based instrumentation} and \textit{replica-based instrumentation}. \textit{Patch-based instrumentation} replaces the instruction with an unconditional branch to a new section containing instrumentation, replaced instruction, and a control flow transfer back to the patch point. Detouring is a direct rewriting and is ISA dependant which makes the approach inconvenient. This approach introduces a high performance degradation given the two control transfers at patch points. 

\textit{Replica-based instrumentation} method inserts jump instructions to a replicated code section containing both a copy of the original code and the instrumentation. All memory references in this section are modified to maintain fewer control flow transfers between original and replicated section. While the performance of this approach is better compared to the \textit{patch-based} instrumentation, the size of the resulting binary is noticeably increased. 

\textit{Reassembleable disassembly} works by recovering relocatable assembly code, the instrumentation of which could be inlined and reassembled back to a working binary. This approach first introduced by UROBOROS \cite{wang2016uroboros} and then expanded by improving on top of the idea in Ramblr \cite{wang2017ramblr}. This approach enhances the performance since inlined assembly avoids inserting control flow changing instructions at instrumentation points. As a result, performance penalty caused by jump instructions are alleviated in this technique. 

\textit{Full-translation} approach works upon translating a low-level machine code to a high-level intermediate representation (IR) using a compiler-based front-end for architecture independent binary rewriting. This process is called \textit{lifting} the binary and assembling the IR back to a working executable is denoted as \textit{lowering}. The advantage of lifting the binary to a high-level IR are two fold. First, relying on IR makes the rewriting framework ISA-agnostic; 
Second, working on a high-level IR provides the ability to apply program analysis techniques like Value Set Analysis (VSA) \cite{balakrishnan2004analyzing} and optimization passes like Simple Expression Tracker (SET) and Offset Shifted Register Analysis (OSRA) \cite{di2017rev}. On the other hand, complete translation suffers from changing the structural integrity such as cache locality and Control Flow Graph (CFG).

While each approach has its own drawbacks and benefits, we focused our evaluation on two recent research: Datalog Disassembly (Ddisasm) \cite{flores2020datalog} for reassembling the disassembly and Rev.ng \cite{di2017rev} as the candidate for full-translation.


\subsection{Comparison of Binary Rewriters}
In this section, we briefly describe the reasons behind choosing the above-mentioned two binary rewriters as our candidates. During the linking phase, linker replaces the symbolic labels with concrete memory addresses which results in losing the relocation information. Hence, to perform rewriting tasks, we need to recover the symbolic references from absolute addresses. A process which is called \textit{symbolization}. Symbolization aims to distinguish whether an intermediate value belongs to a symbol or treat the value as a constant integer. 

Comparing reassembling methods, Ramblr provided counter-examples in real-world binaries for which the UROBOROS symbol categorization fails. UROBOROS scans the data section linearly and considers any machine word-sized buffer whose integer representation falls in a memory region as a memory reference. This assumption with the compiler optimization introduces False-positive and False-negatives. 

Ramblr improved the \textit{content classification} by applying strong heuristics like \textit{localized VSA} and \textit{Intra-function data dependence analysis}. To improve the binary rewriting results, Ramblr depends heavily on symbolic execution for accurate CFG recovery which slows down the rewriting process and brings up scalability issues. 

Apart from Ramblr's heuristics, Ddisasm incorporated register value analysis as an alternative over traditional VSA. In addition, they introduced Data Access Pattern (DAP) analysis which is a def-use analysis combined with the results of register value analysis for a refined register value inference at any given data access point.

Rev.ng relies on full binary translation by lifting the binary to TCG (the IR used in QEMU \cite{bellard2005qemu}) and para-lifting TCG to LLVM-IR to benefit from more advanced transformation and analysis passes for CFG and function boundary recovery. While frameworks like angr \cite{wang2017angr} use lifting to apply more advanced binary analysis on top of the intermediate-level representation, they do not lower the resulting transformation back to the binary. Moreover, Rev.ng heavily relies on code pointers for identifying function entry points and leverages VSA for a more precise value boundary tracking. 

As the rewriter tools to harden the binary code against fault injection, in this work, we chose Ddisasm and Rev.ng to show the difference between two different rewriting schemes for this purpose.

\section{Countermeasure Insertion Methodology}
\label{sec:approaches}

For complex architectures, like x86-64, it is not straight forward to group the bits in the binary file to form full instructions. Neither is it easy to group the instructions to form basic blocks at this level of abstraction.
Therefore, manipulating the binary file directly is not trivial.
We propose a procedure in which we use an open source disassembler and binary manipulation tools as well as binary lifters to make the binary hardening process more manageable.
\subsection{Rewriting the Binary}
\label{sec:rewriting}

Using the disassemblers and working on the assembly code, compared to the binary file, can help in finding patterns of instructions and applying fixes locally. However, at the assembly code level, the register allocation and memory usage are fixed. Therefore, applying fixes at this level requires extra caution not to overwrite the allocated registers in use. A favorable property of this level of hierarchy is that we know which part of the assembly code corresponds to which part of the machine code exactly. We take advantage of this property in our proposed methodology and build an iterative process that, using simulation of fault effects, can locally apply countermeasures only to the parts that they are required. 

While simple and small fixes can be applied at the assembly level, more complex fixes are not easily applicable. In this case, a higher level of abstraction that enables modification of the code and different types of analysis is useful. 
Since LLVM-IR is in Single Static Assignment (SSA) \cite{rosen1988global} format and supports different levels of hierarchy (namely module, function, basic-block, and instruction), we choose it as our high-level IR. Support of different levels of hierarchy in the IR makes it easier to perform static analyses on the program.
Additionally, being a part of the LLVM tool-chain, has the advantage of being open-source, having a big number of active contributors, and a well-maintained documentation. Despite the aforementioned advantages, however, lifting the binary to such a high level of abstraction will eliminate the mapping between the abstraction levels. This results from the fact that the high level of abstraction lacks the low-level target-dependent information.
Consequently, applying targeted and local fixes to the binary files at this high level of abstraction is not readily available.

In the following subsections, we discuss the mentioned approaches to countermeasure injection.

\subsection{Faulter+Patcher Approach}
\label{sec:binarycntr}

\begin{figure}
    \centering
    \includegraphics[width=.7\linewidth]{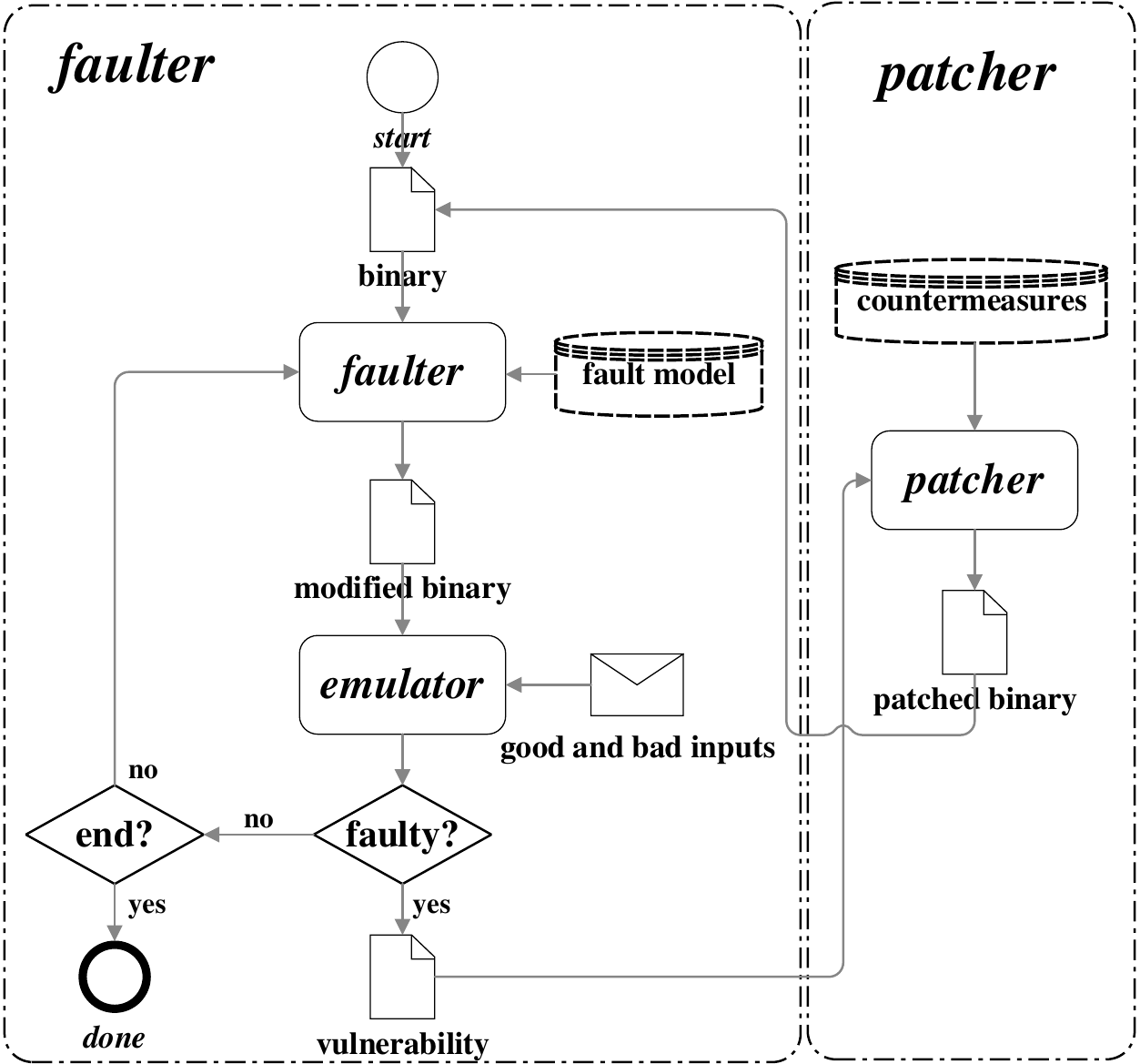}
    \caption{Flowchart of the \textit{Faulter+Patcher} approach}
    \label{fig:binaryAproachFlowchart}
    \vspace{-.6cm}
\end{figure}

Our first approach injects countermeasures at the abstraction level of assembly code and is thus able to patch the binary file in a targeted manner. Fig.~\ref{fig:binaryAproachFlowchart} shows the overall scheme. In this approach, we have a fault simulation-driven, iterative method to mitigate fault injection vulnerabilities in the binary file. The system consists of two main components: a \textit{faulter} and a \textit{patcher}. The \textit{faulter} is simulating faults under a certain fault model in a target binary and produces a list of vulnerabilities,
meaning faults where unwanted behavior in the target binary is triggered. The \textit{patcher} uses the list produced by the \textit{faulter} to patch the binary. The \textit{patcher} will patch each fault vulnerability as localized as possible, without affecting its surrounding code. The patched binary is then again run through the \textit{faulter} and \textit{patcher}. We repeat this process until no more faults are present or can be fixed. In the following, the \textit{faulter} and the \textit{patcher} are discussed in detail.

\subsubsection{Faulter}
For our purpose, fault injection vulnerabilities are vulnerabilities where an attacker, i.e. an unauthorized user, is able to trigger a behavior in a target binary that should be reserved only for authorized users.
For example, consider a pincode checker that receives an input pin and checks if the inserted value is correct. For a correct pin, the program will proceed to run some sensitive operations. An attacker does not know the correct pincode, but may be able to skip an instruction in the target binary such that the program will conclude the inserted pin is correct and therefore run the sensitive operations. These faults are labeled ``successful faults''. Faults that do not trigger the unwanted behavior or cause the program to crash are ignored. 
We first choose a fault model that we want to protect our binary file against. Regardless of the fault model, and the number of faults injected per run, the \textit{faulter} takes a target binary and two inputs; a ``good'' input and a ``bad'' input. 
For instance, in the pincheck example, the ``good'' input is the correct pincode and the ``bad'' input is any value other than the correct pincode.
First, the ``good'' and the ``bad'' inputs are executed to see the difference of execution traces between them. 
When running the target binary with the ``bad'' input, we effectively can record a trace of all the instructions executed. For each offset in that trace, taking the ``single bit flip model'' as an example, we run the target binary normally up to that offset in the trace, flip a bit in the instruction at the trace offset, and then resume execution. The target binary either crashes, executes as an incorrect input, or behaves differently (as a correct input). 
If it behaves as a correct input, the trace offset and the fault that caused it, in this case a bit offset into an instruction at the trace offset, is recorded.

We implemented this simple \textit{faulter} in Python using the Qiling binary emulator package.
We fork each fault simulation to speed up the process. Our \textit{faulter} supports x86-64 Linux binaries only, but including support for other architectures supported by Unicorn should be straightforward. 

\subsubsection{Patcher}
The list of ``successfult faults'' coming out of the \textit{faulter} is addressed locally in the \textit{patcher}. The \textit{patcher} replaces the vulnerable patterns of instructions with known hardened patterns.
For example, consider a run of the \textit{faulter} under the ``instruction skip'' fault model that identified that at timestamp 40, the skipping of a {\tt mov} instruction is a successful fault. A local countermeasure is to perform the {\tt mov} twice, or add a compare instruction to verify the {\tt mov} has been executed prior. Note that these countermeasures cause duplicate reads, as redundancy is key to mitigate fault injection attacks. 


We implemented a proof of concept \textit{patcher} based on GrammaTech's Ddisasm tool and their Python binary manipulation libraries. The Ddisasm tool performs pointer analysis on an executable and produces an IR in the form of GrammaTech Intermediate Representation for Binaries (GTIRB) \cite{schulte2019gtirb} that can then be manipulated and recompiled into an executable by the Python GTIRB libraries. 

\subsubsection{Rinse and repeat}
After running the \textit{faulter} and the \textit{patcher} once, we end up with a patched binary. Running the \textit{faulter} on the patched binary may reveal that, since we added code and changed distances between instructions, we added new vulnerabilities. These new vulnerabilities then can be addressed by running the \textit{patcher} iteratively until a fixed point is reached.

\subsection{Hybrid Compiler-Binary Approach}
\label{sec:hybrid}

In our second approach, we inject countermeasures at the abstraction level of compiler IR.
To be able to implement more complex countermeasures, working at the level of assembly code is cumbersome if not impractical. For example, consider a countermeasure where extra registers are needed to hold some intermediate values. If the assembly code had been generated with a high level of optimization and no register is available in the unprotected program, extra steps should be taken to spill some data to the memory to make a few registers available and load them back to the same registers after the countermeasure. This requires knowledge of the state of the memory at different locations. However, there is no guarantee that these steps are possible therefore implementing some countermeasures on certain programs might not be feasible. 

\begin{figure}
    \centering
    \includegraphics[width=.9\linewidth]{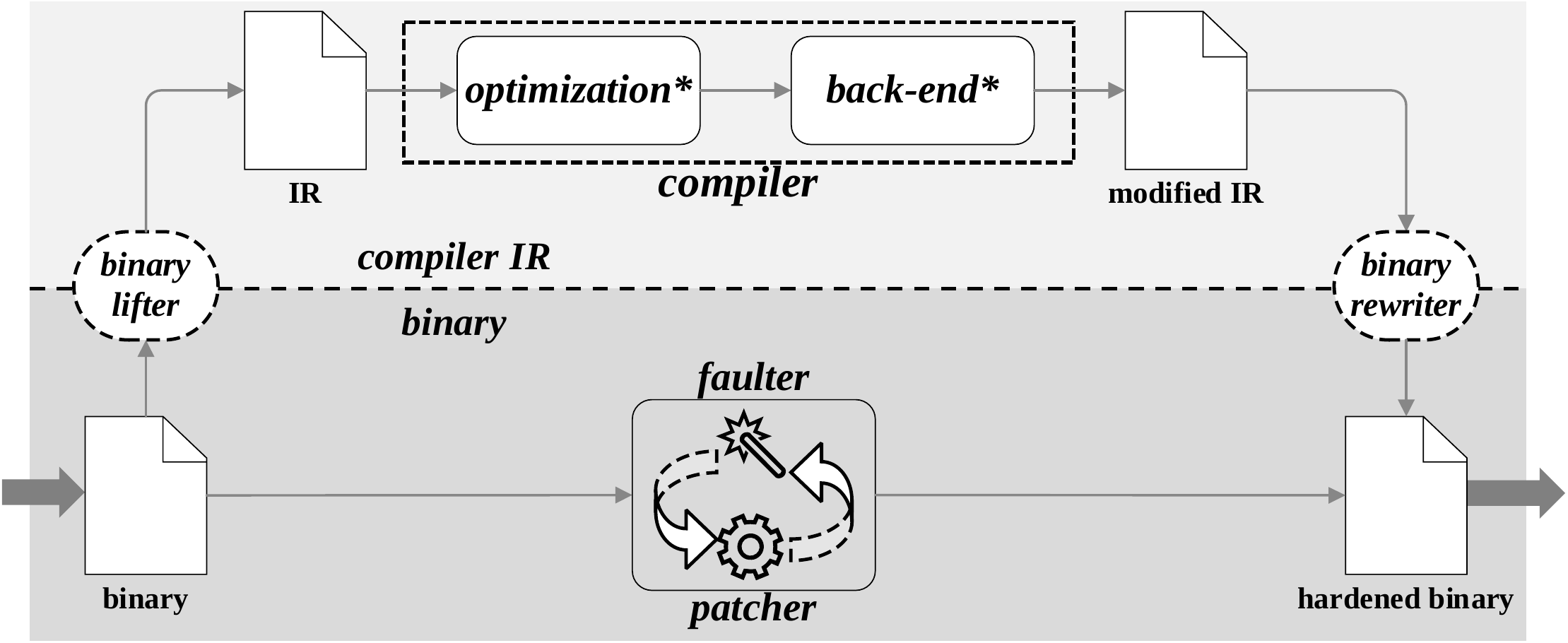}
    \caption{High-level overview of the \textit{Faulter+Patcher} (lower half) and the \textit{Hybrid} (upper half) approaches}
    \label{fig:comparisonApproaches}
    \vspace{-.5cm}
\end{figure}

To overcome this problem, we propose a process as shown in the upper half of Fig.~\ref{fig:comparisonApproaches}.
This process consists of three steps; First, we transform the program from binary to a compiler IR. Second, we implement the countermeasure on the IR. Last, we transform the protected IR back to the executable binary format, hence, achieve the goal of protecting the binary file. These steps are elaborated in the rest if this section. 


\subsubsection{Transforming to an intermediate representation} The goal of this step is to have a representation of the binary file which, while preserving the functionality, is easier to modify and supports a format in which different types of analysis can be performed. As discussed earlier, we choose LLVM-IR.

There are several open-source tools that are able to lift the binary file of different architectures to the architecture-independent form of LLVM-IR. 
In this work, we use Rev.ng. As mentioned in Section~\ref{sec:binarylifter}, Rev.ng is a binary analysis framework based on QEMU and LLVM. As part of this framework, it is possible to extract the LLVM-IR of a program from its binary. The binary file can be for any of the x86, x86-64, ARM, MIPS, s390x, or AArch64 (WIP) architectures. In this project, without loss of generality, our focus is on the x86-64 architecture.

\subsubsection{Implementing the countermeasure} After the lifting step is performed, we have the LLVM-IR representation of the algorithm. Therefore the countermeasure can be implemented in the form of a combination of optimization and/or back-end passes depending on the desired protection. If the protection algorithm does not have any dependencies specific to the target architecture, the entire countermeasure can be in the form of an optimization pass. Otherwise, back-end passes would be required. 

\subsubsection{Generating the protected executable} Finally, the protected IR needs to be compiled to an executable. In LLVM, the {\tt llc} tool is responsible for translating the LLVM-IR to an architecture-specific executable file. Specific steps might need to be added in the form of back-end passes to make sure the implemented countermeasures are retained unchanged through this process. Once the hardened binary file is generated, we use the same \textit{faulter} system to detect remaining vulnerabilities.

\subsection{Choosing the Right Method}

The targeted insertion of countermeasures in the \textit{Faulter+Patcher} scheme makes the overhead of the applied assurance smaller than a holistic approach. Furthermore, the mere act of lifting the binary to LLVM-IR and translating it back to binary in the \textit{Hybrid} approach adds extra overhead to the program. This stems from the internal functions of the binary rewriting tools, which in our case is Rev.ng.
On the other hand, applying countermeasures in the \textit{Hybrid} approach is easily automated and is guaranteed to be feasible. The hierarchy levels supported by LLVM-IR as well as its SSA format eases many static analyses such as finding idempotent pieces, finding and replacing all the uses of a variable, and many more. 

The aforementioned trade-off between these methods, makes each method suitable for a different use case scenario. In size-constrained applications, such as programs for small embedded systems, the \textit{Faulter+Patcher} method is more favorable due to its smaller footprint. In scenarios where the code size is not of critical concern, the \textit{Hybrid} approach provides a simpler and guaranteed assurance for inserting complex countermeasures.





\section{Experimental Results} \label{sec:experiments}

In this section, we show how our proposed approaches can apply countermeasures against a chosen fault model. We first show the local protections that we add by our \textit{Faulter+Patcher} approach. We then demonstrate a holistic protection that can be used in our \textit{Hybrid} approach. Finally, we show the results of the inserted countermeasures in our case studies.

\subsection{Local Protections}
\label{sec:localcntr}

\begin{table}[t]
\caption{Local protection pattern for {\tt mov} operations}
\centering
\begin{tabular}{ll}
\hline
\multicolumn{1}{c}{\textbf{Original}}                                         & \multicolumn{1}{c}{\textbf{Protected}}                                                                                                                                                 \\ \hline
\begin{tabular}[c]{@{}l@{}}mov rax, {[}rbx+4{]}\\ happyflow: ...\end{tabular} & \begin{tabular}[c]{@{}l@{}}mov rax, {[}rbx+4{]}\\ cmp rax, {[}rbx+4{]}\\ je happyflow\\ call faulthandler\\ happyflow: ...\end{tabular} \\ \hline
\end{tabular}
\vspace{-.5cm}
\label{tab:mov}
\end{table}



\begin{table}[t]
\caption{Local protection pattern for {\tt cmp} operations}
\centering
\scalebox{0.85}{

\begin{tabular}{ll}
\hline
\multicolumn{1}{c}{\textbf{Original}}                                           & \textbf{Protected}                                                                                                                                                                                                                                          \\ \hline
\begin{tabular}[c]{@{}l@{}}cmp rbx, {[}rcx+4{]}\\ fallthrough: ...\end{tabular} & \begin{tabular}[c]{@{}l@{}}lea rsp, {[}rsp-128{]}\\ cmp rbx, {[}rcx+4{]}\\ push rbx\\ pushfq\\ cmp rbx, {[}rcx+4{]}\\ pushfq\\ pop rbx\\ cmp rbx, {[}rsp{]}\\ je restore\\ call faulthandler\\ resotre:\\ popfq\\ pop rbx\\ lea rsp, {[}rsp+128{]}\\ fallthrough: ...\end{tabular} \\ \hline
\end{tabular}
}
\label{tab:cmp}
\end{table}
\begin{table}[b]
\caption{Local protection for conditional jump operation}
\centering
\scalebox{0.8}{

\begin{tabular}{ll}
\hline
\multicolumn{1}{c}{\textbf{Original}}                                                                                & \multicolumn{1}{c}{\textbf{Protected}}                                                                                                                                                                                                                                                       \\ \hline
\begin{tabular}[c]{@{}l@{}}j\textless{}cond\textgreater jumptarget\\ fallthrough: ...\\ jumptarget: ...\end{tabular} & \begin{tabular}[c]{@{}l@{}}j\textless{}cond\textgreater newjumptarget\\ lea rsp, {[}rsp-128{]}\\ push rcx\\ pushfq\\ set cl\\ cmp cl, 0\\ je newfallthroughjmp\\ call faulthandler\\ newfallthroughjmp:\\ popfq\\ pop rcx\\ j\textless{}cond\textgreater fallthrough\\ call faulthandler\\ newjumptarget:\\ lea rsp, {[}rsp-128{]}\\ push rcx\\ pushfq\\ set cl\\ cmp cl, 1\\ je newjumptargetjmp\\ call faulthandler\\ newjumptargetjmp:\\ popfq\\ pop rcx\\ j\textless{}cond\textgreater jumptarget\\ call faulthandler\\ fallthrough: ...\\ jumptarget: ...\end{tabular} \\ \hline
\end{tabular}
}
\label{tab:jmp}
\end{table}

In the \textit{Faulter+Patcher} approach, we are able to insert protected code patterns locally. The following is the description of these redundant computation-based protections.

\subsubsection{Protecting {\tt mov} Instruction} To protect the {\tt mov} operation against fault attacks, after executing the {\tt mov} operation, the result of the two memory locations are compared and in case of an inconsistency, a faulthandler is called (Table~\ref{tab:mov}).

\subsubsection{Protecting the {\tt cmp} Instruction}
We can protect a {\tt cmp} instruction against fault attacks by executing the comparison twice and comparing their resulting flags (Table~\ref{tab:cmp}). To this end, we use the {\tt pushfq} instruction in x86-64 ISA which requires a valid stack pointer ({\tt rsp}). Due to Intel's red zone, we have to subtract 128 bytes from {\tt rsp} to jump out of the red zone.

\subsubsection{Protecting the {\tt j<cond>} operation}
By hardening the conditional jump operations, we detect glitches that change the jump condition. In the protected code shown in Table~\ref{tab:jmp}, we use the flags register and match this to the expected flag in the jump-target and the fall-through of a branch.

\subsection{Holistic Protection} 
\label{sec:holisticcntr}

\begin{verbbox}[\small]
        cmp rs1, rs2
        bne target2
        ...
    target1:
        ...
    target2: 
        ...
\end{verbbox}
\begin{figure}[b]
  \centering
  \theverbbox\qquad\includegraphics[width=.4\linewidth]{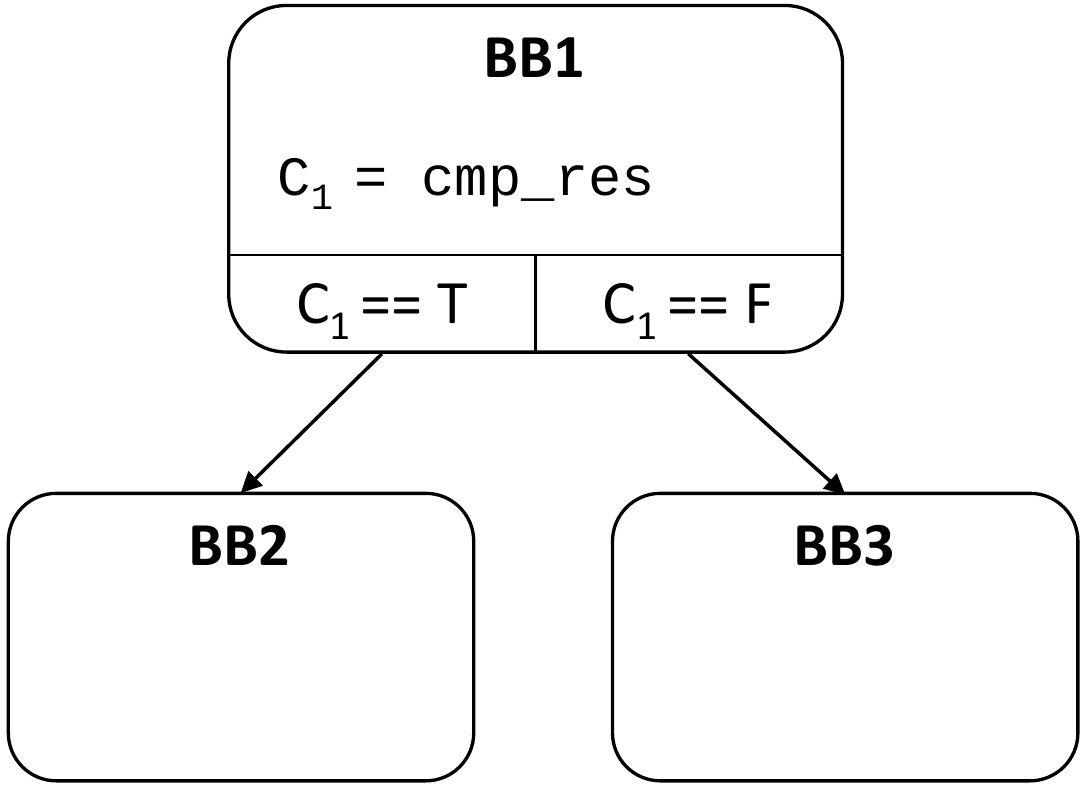}
    \caption{Assembly code and CFG of a simple branch instruction}
    \label{fig:branchInst}
\end{figure}

In this section, we describe a conditional branch hardening method that we later use in our case studies for the \textit{Hybrid} countermeasure implementation approach.

Imagine a simple program that receives a pin code and only if the pin code is correct, resumes the program to execute some operations. In the assembly code (equally the executable file) of this program, there will be a comparison instruction to compare the inserted pin code and its expected value, as well as a conditional branch that, based on the result of the comparison, jumps to a successor basic block. Figure~\ref{fig:branchInst} shows the assembly code and the control flow graph (CFG) of this branch operation.
In the case of Control Flow Integrity (CFI), going from BB1 to either of BB2 or BB3 does not raise any issues since they both are valid paths in the CFG. However, if an injected fault results in taking the wrong branch, it will be an unnoticed fault.  
In this conditional branch hardening method, our goal is to protect against this outcome of FI. 

In this method, we assign a unique ID to each basic block ($UID_{BB}$) at compile time. 
We then use an algorithm, \(h(UID_{src},UID_{dst},cmp\_res)\), which calculates a checksum at run-time based on the fixed $UID$s assigned to the source and destination blocks at compile-time, i.e. $UID_{src}$ and $UID_{dst}$, and the dynamically-evaluated compare result, $cmp\_res$. 
The calculated checksum will be stored in a register and checked in the destination basic blocks. At the destination blocks, since the expected $cmp\_res$ for the taken edge is known, the expected value of $h$ is known. Therefore checking the evaluated value only requires reading the register value and comparing it with the expected value. When the register does not contain the expected value, we jump to a fault-response basic block.

The simplicity level of the $h$ function can be decided based on the required security properties of the program. 
As an example, we chose a simple option for $h$ and implemented the countermeasure as an optimization pass in the LLVM tool-chain. In this example, the checksum is calculated as the XOR result of the UID of the taken destination block and that of the source block ($UID_{dst} \oplus UID_{src}$).
The pseudo-code of the calculation procedure of this checksum in LLVM is shown in Algorithm~\ref{alg:h_alg} where $cmp\_res$ is the result of the comparison for the conditional branch, and $UID_{Tdst}$, $UID_{Fdst}$, and $UID_{src}$ are the UIDs assigned to the true destination (the destination taken when the comparison result is true), the false destination (the destination taken when the comparison result is false), and the source block respectively. The $mask$ shown on line 4 will have the value of all ones if the comparison result is false and the value of all zeros if the comparison result is true. The checksum will be located in one register and each of the destination blocks will evaluate whether the value of the checksum is correct. 

\begin{figure}[b]
    \centering
    \includegraphics[width=.7\linewidth]{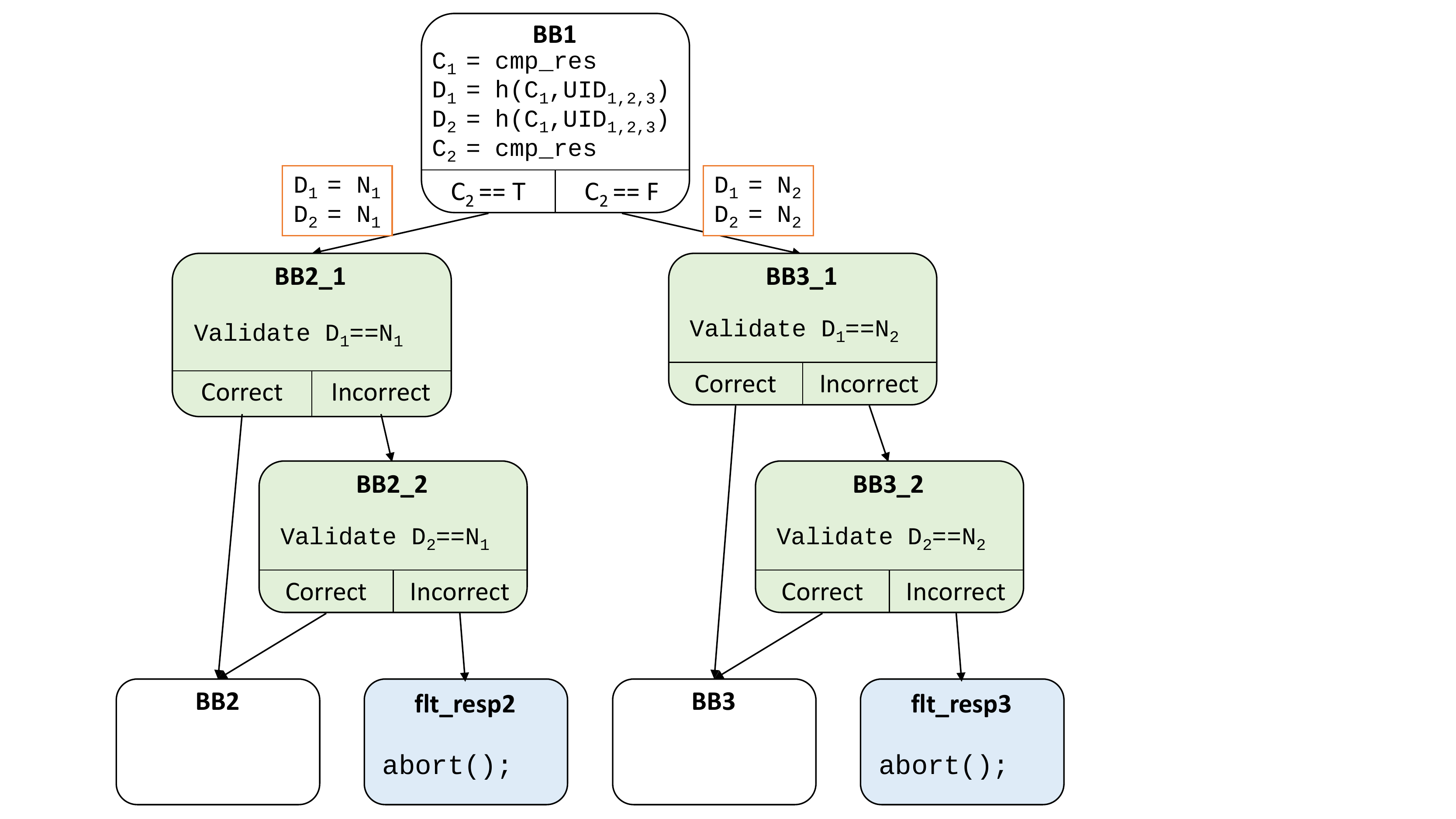}
    \caption{CFG of the example conditional branch hardening}
    \label{fig:countermeasure_cfg}
\end{figure}

Furthermore, we made this evaluation duplicated; 
Figure~\ref{fig:countermeasure_cfg} shows the CFG of this implementation protecting the conditional branch shown in Figure~\ref{fig:branchInst}. We run the comparison instruction once ({\tt C\textsubscript{1}}), and calculate the checksum based on its result and keep it in a register ({\tt D\textsubscript{1}}). We perform this calculation another duplicated time and keep the result in a new register ({\tt D\textsubscript{2}}). We then perform the comparison again and run the branch based on the result of the second comparison ({\tt C\textsubscript{2}}). As shown by the orange boxes, the expected value of the checksum is different for the out-going edges of the source block ({\tt N\textsubscript{1}} vs. {\tt N\textsubscript{2}}). 
In the destination blocks, we check both copies of the checksum stored in registers against the expected values in a nested fashion. The green blocks show the nested checksum validations and the blue blocks represent the fault-response. As a simple example, the fault-response can be aborting the execution (in the {\tt flt\_respx} basic blocks).

 \begin{algorithm}[t]

     \small
 \caption{Simple example algorithm for $h$}
 \label{alg:h_alg}
 
 \begin{algorithmic}[1]
 \renewcommand{\algorithmicrequire}{\textbf{Input:}}
 \renewcommand{\algorithmicensure}{\textbf{Output:}}
 \REQUIRE $cmp\_res$, $UID_{Tdst}$, $UID_{Fdst}$, $UID_{src}$
 \ENSURE $checksum$
 \\ \textit{Generate unique checksums for edges} :
 \STATE $const_{Tdst} \gets UID_{Tdst} \oplus UID_{src}$
 \\ \STATE $const_{Fdst} \gets UID_{Fdst} \oplus UID_{src}$
 \\ \textit{Calculate checksum of the branch} :
  \STATE $cmp\_ext \gets zero\_extend(cmp\_res)$
  \\ \STATE $mask \gets cmp\_ext - 1$
  \\ \STATE $checksum \gets (\neg mask \land const_{Tdst}) \lor (mask \land const_{Fdst})$
 \end{algorithmic}
 \vspace{-.6cm}
 \end{algorithm}

In this scenario, if the attacker tries to skip one of the comparison instructions or change it to compute the inverse output, the checksum validation process will be able to catch the injected error. If the adversary intends to bypass this process, they would need to inject the exact same fault in both comparison results. In Section~\ref{sec:casestudies}, we show a simulated analysis of the effectiveness of this countermeasure.

The overall overhead of this countermeasure depends on the number of conditional branches that we want to protect and therefore is highly application-specific. Table~\ref{tab:qualOverhead} shows how many instructions are required to replace a simple conditional branch by this method. We implement the conditional branch hardening as an optimization pass in the LLVM compiler tool-chain. Therefore, its translation will differ for different target architectures based on how the instruction-lowering is done for that architecture in the LLVM back-end. 

\begin{table}[t]
\caption{Qualitative overhead of the conditional branch hardening}
\centering
\scalebox{0.85}{

\begin{tabular}{c|c|c|c}
\hline
\multicolumn{2}{c|}{\textbf{Before Protection}}                                                                          & \multicolumn{2}{c}{\textbf{After Protection}}                                                                                                                                                                                                    \\
\textbf{LLVM-IR}                                     & \textbf{x86-64}                                                   & \textbf{LLVM-IR}                                                                                        & \textbf{x86-64}                                                                                                                         \\ \hline
\begin{tabular}[c]{@{}c@{}}1 cmp\\ 1 br\end{tabular} & \begin{tabular}[c]{@{}c@{}}1 cmp\\ 1 jx (cond. jump)\end{tabular} & \begin{tabular}[c]{@{}c@{}}1 cmp\\ 2 zext\\ 2 sub\\ 6 xor\\ 2 or\\ 4 and\\ 1 br\\ 4 switch\end{tabular} & \begin{tabular}[c]{@{}c@{}}2 cmp\\ 6 mov\\ 2 sub\\ 6 xor\\ 2 or\\ 6 and\\ 2 test\\ 4 jx (cond. jump)\\ 5 jmp (uncond. jump)\end{tabular} \\ \hline
\end{tabular}

}
\label{tab:qualOverhead}
\end{table}

\subsection{Case Studies}\label{sec:casestudies}
We choose two applications for the proof of concept of our proposed methods. The first application is a simple pin-check program that receives an input password and checks the correctness of the inserted password. The second application is a secure bootloader in which the hash of the content of a memory location is calculated and compared with an expected hash value.

For each case study, we use the \textit{faulter} described in Section~\ref{sec:binarycntr} with the two fault models of ``instruction skip'' and ``single bit flip'' and verify that the code has vulnerabilities to these fault models. All of these vulnerabilities were caused by the conditional jumps ({\tt mov}, {\tt cmp}, and {\tt jmp} instructions related to a jump operation) in the program.
We then insert the code patterns described in Section~\ref{sec:localcntr} using the \textit{Faulter+Patcher} approach for each vulnerable point and reevaluate the hardened binary file iteratively until confirmed that no more vulnerabilities exist. Furthermore, we apply the conditional branch hardening countermeasure from Section~\ref{sec:holisticcntr} using our \textit{Hybrid} approach and verify that the vulnerabilities have been mitigated. In the case of the ``instruction skip'' fault model, we were able to resolve all the vulnerabilities using the mentioned countermeasures. In the case of the ``single bit flip'' fault model we were able to reduce the number of vulnerable points by 50\% using both methodologies.  

Table~\ref{tab:casestudyOverhead} shows the overhead caused by the inserted countermeasure in each approach. The overhead caused by the \textit{Hybrid} approach is 2 to 5 times bigger than that of the iterative method. This is an expected outcome since the \textit{Hybrid} approach applies the countermeasure to the entire program whereas the \textit{Faulter+Patcher} approach only does so to the vulnerable points. Furthermore, duplicating every instruction, which is the go-to protection scheme against fault injection, implies at least 300\% overhead in code size (since for each instruction, it will add another copy of the instruction and a comparison procedure between their results). Therefore, both of our methods perform better than a simple duplication scheme.



\begin{table}[t]
\caption{Overhead of adding the protections}
\centering

\begin{tabular}{l|cc}
\hline
\multirow{2}{*}{\textbf{Case Study}} & \multicolumn{2}{c}{\textbf{Overhead in code size (\%)}} \\
                                     & \textbf{\textit{Faulter+Patcher}} & \textbf{Hybrid} \\ \hline
pincheck                             & 17.61                     & 85.88           \\ 
secure bootloader                    & 19.67                     & 48.67           \\ \hline
\end{tabular}

\label{tab:casestudyOverhead}
\end{table}

%


%

\section{Conclusion and Future Work}\label{sec:conclusion}
In this work, we proposed two approaches for hardening the binary file against fault attacks based on binary rewriting. In the first approach, we disassemble the binary file and insert countermeasures which enables us to insert local countermeasure and keep the structure of the original binary file. We propose an iterative and simulation-driven framework that only inserts countermeasures to the vulnerable parts of the program. In the second approach, we lift the binary file to LLVM-IR and insert countermeasures at a higher level of abstraction. This enables us to implement more complex countermeasures and perform static analyses on the program.

As the future work, we intend to study the mapping between the lifted LLVM-IR of a binary code and the original binary file and enable an iterative countermeasure insertion for the \textit{Hybrid} methodology. 



%





\ifCLASSOPTIONcaptionsoff
  \newpage
\fi




\bibliographystyle{IEEEtran}
\bibliography{Bibliography}

\begin{thebibliography}{10}
\providecommand{\url}[1]{#1}
\csname url@rmstyle\endcsname
\providecommand{\newblock}{\relax}
\providecommand{\bibinfo}[2]{#2}
\providecommand\BIBentrySTDinterwordspacing{\spaceskip=0pt\relax}
\providecommand\BIBentryALTinterwordstretchfactor{4}
\providecommand\BIBentryALTinterwordspacing{\spaceskip=\fontdimen2\font plus
\BIBentryALTinterwordstretchfactor\fontdimen3\font minus
  \fontdimen4\font\relax}
\providecommand\BIBforeignlanguage[2]{{%
\expandafter\ifx\csname l@#1\endcsname\relax
\typeout{** WARNING: IEEEtran.bst: No hyphenation pattern has been}%
\typeout{** loaded for the language `#1'. Using the pattern for}%
\typeout{** the default language instead.}%
\else
\language=\csname l@#1\endcsname
\fi
#2}}

\bibitem{piret2003differential}
G.~Piret and J.-J. Quisquater, ``A differential fault attack technique against
  spn structures, with application to the aes and khazad,'' in
  \emph{International workshop on cryptographic hardware and embedded
  systems}.\hskip 1em plus 0.5em minus 0.4em\relax Springer, 2003, pp. 77--88.

\bibitem{leveugle2009statistical}
R.~Leveugle, A.~Calvez, P.~Maistri, and P.~Vanhauwaert, ``Statistical fault
  injection: Quantified error and confidence,'' in \emph{2009 Design,
  Automation \& Test in Europe Conference \& Exhibition}.\hskip 1em plus 0.5em
  minus 0.4em\relax IEEE, 2009, pp. 502--506.

\bibitem{7774479}
N.~{Timmers}, A.~{Spruyt}, and M.~{Witteman}, ``Controlling pc on arm using
  fault injection,'' in \emph{2016 Workshop on Fault Diagnosis and Tolerance in
  Cryptography (FDTC)}, 2016, pp. 25--35.

\bibitem{8419264}
A.~{Vasselle}, H.~{Thiebeauld}, Q.~{Maouhoub}, A.~{Morisset}, and
  S.~{Ermeneux}, ``Laser-induced fault injection on smartphone bypassing the
  secure boot-extended version,'' \emph{IEEE Transactions on Computers},
  vol.~69, no.~10, pp. 1449--1459, 2020.

\bibitem{belleville2018automatic}
N.~Belleville, K.~Heydemann, D.~Courouss{\'e}, T.~Barry, B.~Robisson,
  A.~Seriai, and H.-P. Charles, ``Automatic application of software
  countermeasures against physical attacks,'' in \emph{Cyber-Physical Systems
  Security}.\hskip 1em plus 0.5em minus 0.4em\relax Springer, 2018, pp.
  135--155.

\bibitem{lalande2014software}
J.-F. Lalande, K.~Heydemann, and P.~Berthom{\'e}, ``Software countermeasures
  for control flow integrity of smart card c codes,'' in \emph{European
  Symposium on Research in Computer Security}.\hskip 1em plus 0.5em minus
  0.4em\relax Springer, 2014, pp. 200--218.

\bibitem{barry2016compilation}
T.~Barry, D.~Courouss{\'e}, and B.~Robisson, ``Compilation of a countermeasure
  against instruction-skip fault attacks,'' in \emph{Proceedings of the Third
  Workshop on Cryptography and Security in Computing Systems}, 2016, pp. 1--6.

\bibitem{given2019automated}
T.~Given-Wilson, A.~Heuser, N.~Jafri, and A.~Legay, ``An automated and scalable
  formal process for detecting fault injection vulnerabilities in binaries,''
  \emph{Concurrency and Computation: Practice and Experience}, vol.~31, no.~23,
  p. e4794, 2019.

\bibitem{brejon2019fault}
J.-B. Br{\'e}jon, K.~Heydemann, E.~Encrenaz, Q.~Meunier, and S.-T. Vu, ``Fault
  attack vulnerability assessment of binary code,'' in \emph{Proceedings of the
  Sixth Workshop on Cryptography and Security in Computing Systems}, 2019, pp.
  13--18.

\bibitem{de2016link}
R.~De~Keulenaer, J.~Maebe, K.~De~Bosschere, and B.~De~Sutter, ``Link-time smart
  card code hardening,'' \emph{International Journal of Information Security},
  vol.~15, no.~2, pp. 111--130, 2016.

\bibitem{van2005diablo}
L.~Van~Put, D.~Chanet, B.~De~Bus, B.~De~Sutter, and K.~De~Bosschere, ``Diablo:
  a reliable, retargetable and extensible link-time rewriting framework,'' in
  \emph{Proceedings of the Fifth IEEE International Symposium on Signal
  Processing and Information Technology, 2005.}\hskip 1em plus 0.5em minus
  0.4em\relax IEEE, 2005, pp. 7--12.

\bibitem{o2011retrofitting}
P.~O’sullivan, K.~Anand, A.~Kotha, M.~Smithson, R.~Barua, and A.~D.
  Keromytis, ``Retrofitting security in cots software with binary rewriting,''
  in \emph{IFIP International Information Security Conference}.\hskip 1em plus
  0.5em minus 0.4em\relax Springer, 2011, pp. 154--172.

\bibitem{anand2010decompilation}
K.~Anand, M.~Smithson, A.~Kotha, K.~Elwazeer, and R.~Barua, ``Decompilation to
  compiler high ir in a binary rewriter,'' \emph{University of Maryland, Tech.
  Rep}, 2010.

\bibitem{wang2016uroboros}
S.~Wang, P.~Wang, and D.~Wu, ``Uroboros: Instrumenting stripped binaries with
  static reassembling,'' in \emph{2016 IEEE 23rd International Conference on
  Software Analysis, Evolution, and Reengineering (SANER)}, vol.~1.\hskip 1em
  plus 0.5em minus 0.4em\relax IEEE, 2016, pp. 236--247.

\bibitem{wang2017ramblr}
R.~Wang, Y.~Shoshitaishvili, A.~Bianchi, A.~Machiry, J.~Grosen, P.~Grosen,
  C.~Kruegel, and G.~Vigna, ``Ramblr: Making reassembly great again.'' in
  \emph{NDSS}, 2017.

\bibitem{balakrishnan2004analyzing}
G.~Balakrishnan and T.~Reps, ``Analyzing memory accesses in x86 executables,''
  in \emph{International conference on compiler construction}.\hskip 1em plus
  0.5em minus 0.4em\relax Springer, 2004, pp. 5--23.

\bibitem{di2017rev}
A.~Di~Federico, M.~Payer, and G.~Agosta, ``rev.ng: a unified binary analysis
  framework to recover {CFG}s and function boundaries,'' in \emph{Proceedings
  of the 26th International Conference on Compiler Construction}, 2017, pp.
  131--141.

\bibitem{flores2020datalog}
A.~Flores-Montoya and E.~Schulte, ``Datalog disassembly,'' in \emph{29th
  $\{$USENIX$\}$ Security Symposium ($\{$USENIX$\}$ Security 20)}, 2020.

\bibitem{bellard2005qemu}
F.~Bellard, ``Qemu, a fast and portable dynamic translator.'' in \emph{USENIX
  Annual Technical Conference, FREENIX Track}, vol.~41, 2005, p.~46.

\bibitem{wang2017angr}
F.~Wang and Y.~Shoshitaishvili, ``Angr-the next generation of binary
  analysis,'' in \emph{2017 IEEE Cybersecurity Development (SecDev)}.\hskip 1em
  plus 0.5em minus 0.4em\relax IEEE, 2017, pp. 8--9.

\bibitem{rosen1988global}
B.~K. Rosen, M.~N. Wegman, and F.~K. Zadeck, ``Global value numbers and
  redundant computations,'' in \emph{Proceedings of the 15th ACM SIGPLAN-SIGACT
  symposium on Principles of programming languages}, 1988, pp. 12--27.

\bibitem{schulte2019gtirb}
E.~Schulte, J.~Dorn, A.~Flores-Montoya, A.~Ballman, and T.~Johnson, ``Gtirb:
  intermediate representation for binaries,'' \emph{arXiv preprint
  arXiv:1907.02859}, 2019.

\end{thebibliography}

\end{document}